\documentclass{article}
\usepackage{spconf,amsmath,graphicx,multirow,tabularx, booktabs,array}

\title{dynamic latency speech recognition with asynchronous revision}
\name{Mingkun Huang, Yang Zhang, Meng Cai, Jun Zhang, Yongbin You, Yi He, Zejun Ma}
\address{Bytedance AI-Lab \\
{\small \tt huangmingkun@bytedance.com}}
\begin{document}
%
\maketitle
\begin{abstract}
In this work we propose an inference technique, \emph{asynchronous revision}, to unify streaming and non-streaming speech recognition models. Specifically, we achieve \emph{dynamic latency} with only one model by using arbitrary right context during inference. The model is composed of a stack of convolutional  layers for audio encoding. In inference stage, the history states of encoder and decoder can be asynchronously revised to trade off between the latency and the accuracy of the model. 
To alleviate training and inference mismatch, we propose a training technique, \emph{segment cropping}, which randomly splits input utterances into several segments with forward connections. This allows us to have dynamic latency speech recognition results with large improvements in accuracy. 
Experiments show that our dynamic latency model with asynchronous revision gives 8\%-14\% relative improvements over the streaming models.
\end{abstract}
\section{Introduction}
\label{sec:intro}


End-to-end (E2E) models \cite{graves2012rnnt,graves2013rnnt,william2016las,kim2017joint,google2018sota-las,gulati2020conformer} gained large popularity for ASR over the last few years. These models, which combine the acoustic, pronunciation and language models into a single network, have shown competitive results compared to conventional ASR systems. 

Recently, there has been considerable interest in training E2E models for streaming ASR \cite{google2017stream_rnnt,google2019streaming_mobile,moritz2020streaming,sainath2020streaming,zhang2020transformer}. It is shown that having access to future context to encode the current frame significantly improves speech recognition accuracy. Bidirectional encoders take advantage of future context, however only full input sequence provided can inference be performed. Therefore, streaming recognition can be achieved by specifying a limited right context, with a cost of accuracy degradation.

Typically, streaming and non-streaming ASR models are usually trained separately. Specially, in streaming condition, various models are trained to reach a suitable latency and accuracy trade-off. In this work, we propose an inference technique, asynchronous revision, to unify streaming and non-streaming speech recognition models. Moreover, in streaming mode, we achieve dynamic latency ASR with only one model. To be specific, the non-streaming model can be used with arbitrary right context during inference. In inference stage, chunk-based incremental decoding is applied on the non-streaming model, the history encoder and decoder states of which can be asynchronously revised to achieve dynamic latency ASR. When performing asynchronous revision during inference,  the encoder  network  may  get  incomplete  right context. To alleviate this training and inference mismatch, we propose a training technique, segment cropping, which randomly splits input utterances into several segments with forward connections. We show that the dynamic latency model gives 8\%-14\% relative improvements over streaming models with the same latency.

\section{related work}
\label{sec:related}
There has been a growing interest in building streaming ASR systems based on RNN-T\cite{graves2012rnnt,graves2013rnnt}.  Compared with attention-based encoder decoder (AED)\cite{william2016las} models, RNN-T models are naturally streamable and have shown great potentials for low-latency streaming ASR. In this work, we mainly focus on RNN-T based models. 

To improve the latency of AED models, partial hypothesis selection is proposed in \cite{liu2020las_retranslation}. We both use chunk-based incremental decoding during inference, the main difference is that we achieve truly streaming ASR based on RNN-T, while they use global attention mechanism. Similar work is explored in simultaneous (speech) translation\cite{arivazhagan2020retranslation,arivazhagan2020retranslation_spoken}. Variable context training is used in Y-model \cite{tripathi2020transformer} architecture. In inference time, the context length for the variable context layers can be changed. However, the latency of Y-model is predefined in training stage. At inference, limited latency configurations can be used. Universal ASR \cite{yu2020universal} is a unified framework to train a single E2E ASR model for both streaming and non-streaming speech recognition. Nevertheless, only one streaming mode is available.

Unlike these approaches, our work not only explores the unification of streaming and non-streaming ASR models, but also achieve dynamic latency ASR with only one model. In other words, with asynchronous revision decoding technique, the non-streaming model can be used with any right context during inference.

\section{dynamic latency asr}
\label{sec:dynamic}

Morden E2E ASR systems have an encoder-decoder structure~\cite{graves2012rnnt,william2016las}. The encoder network encode Mel-spectrogram feature into hidden representation containing semantic information. Then, the decoder network make predictions based on the current output of encoder and history prediction states. In this work, we focus on the most commonly used RNN-T model. 

\subsection{Frame Synchronous Decoding}

\begin{figure}[ht]
  \centering
  \centerline{\includegraphics[width=0.8\columnwidth]{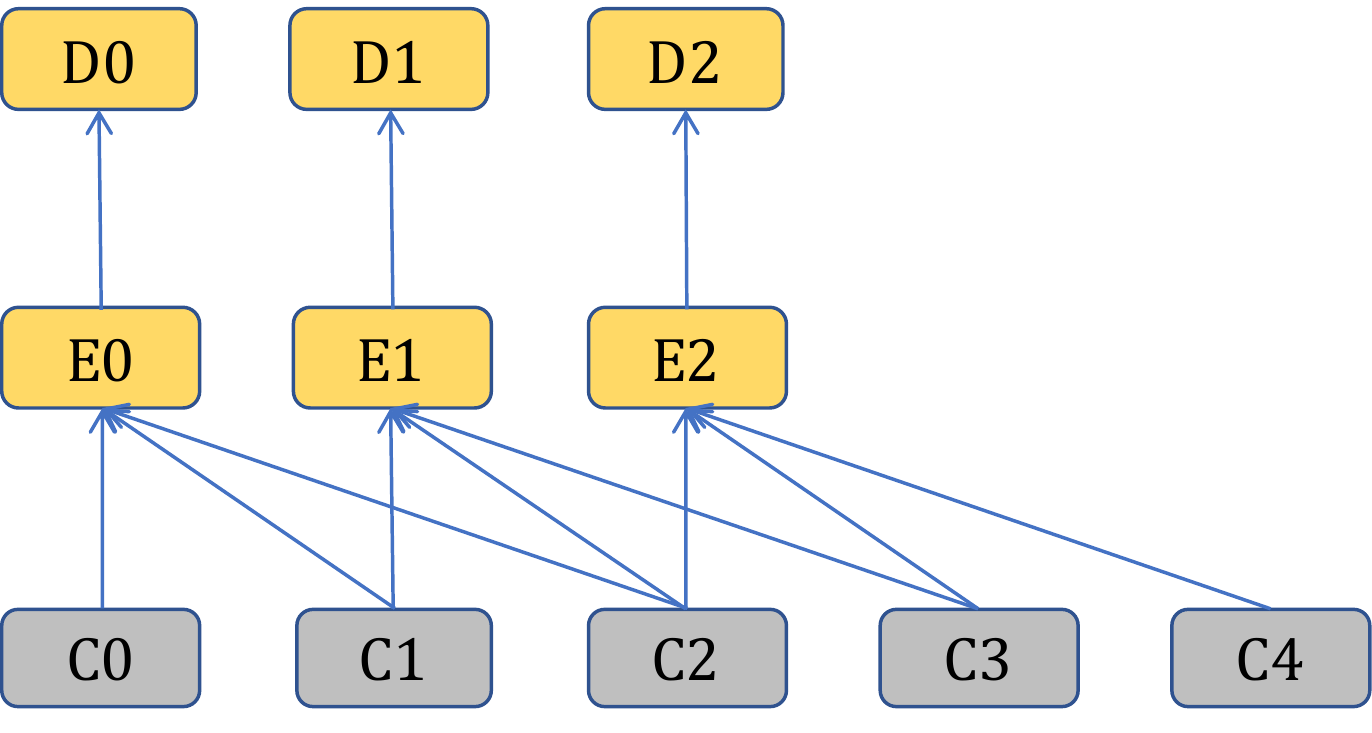}}
\caption{Frame synchronous decoding of RNN-T model. Ci, Ei and Di mean ith input chunk, encoder state and decoder state respectively. Suppose the right context of the encoder network has two chunks, the decoder state is finalized iff complete right context is provided. The incremental dependency of the encoder and decoder states are omitted, along with the left context of the encoder network.}
\label{fig:frame_sync}
\end{figure}

It is clear that the decoder in RNN-T is an auto-regressive model in both streaming and non-streaming ASR models. When performing streaming inference, complete input context for the encoder is necessary. Commonly, for incremental inference, we divide the input utterances into fixed-size chunks and decode every time a new chunk arrives. As shown in figure \ref{fig:frame_sync}, when performing incremental decoding, the encoder state is conditioned on the current chunk and history states (memory). Meanwhile, the decoder state is based on the current encoder state and previous decoder state. 

\subsection{Asynchronous Revision Decoding}
\label{ssec:async}
\begin{figure*}[!ht]
    \centering
    \includegraphics[width=1.0\linewidth]{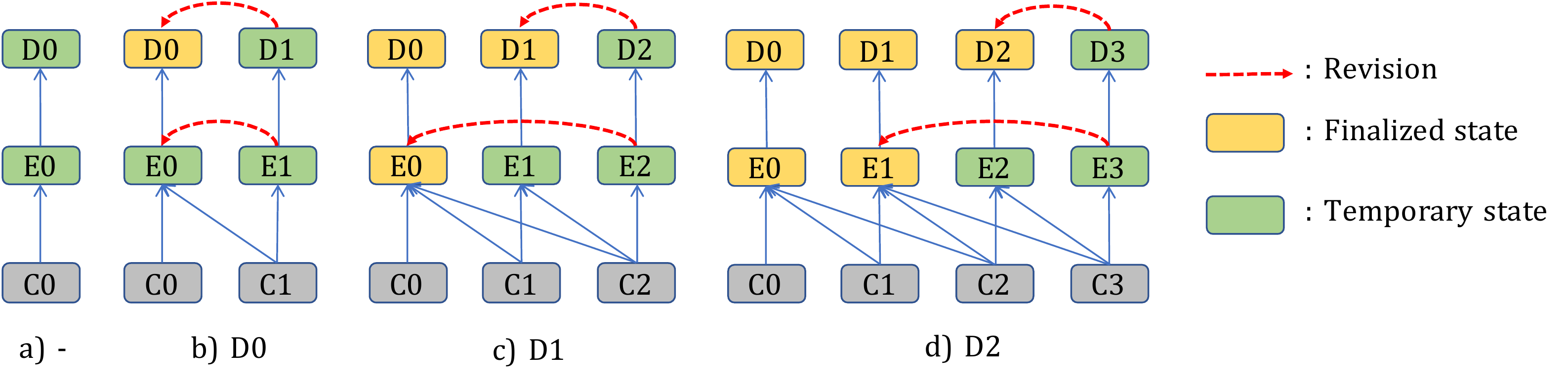}
    \caption{Asynchronous revision decoding of RNN-T model. The right context of the encoder network is limited for non-streaming models when performing incremental decoding. The last several encoder states are not finalized due to incomplete input context. Therefore, revision of the encoder and decoder states is applied for the last one and two steps respectively. At each decoding step, the temporary state will be replaced with the finalized state. The memory cache (forward connection) of the encoder and decoder states are omitted, along with the left context of the encoder network.}
    \label{fig:async_revise}
\end{figure*}

For non-streaming model, its hard to perform incremental decoding since the long right context leads to large latency. What if we limit the right context? Figure \ref{fig:async_revise} depicts the limited context decoding process for non-streaming model. Suppose the right context of the encoder network has two chunks, figure \ref{fig:async_revise} part a shows the initial states when decoding chunk C0. Since the right context of E0 is not complete, when chunk C1 comes, we need to revise the encoder state E0 along with the decoder state D0. As shown in figure \ref{fig:async_revise} part b, if we only revise the last decoder state, then D0 is finalized. After chunk C2 comes, the encoder state E0 get full input context, then we revise E0 to make it finalized as shown in part c. The finalized states will not be changed when new chunks arrive as shown in part d. Temporary states will be revised in the next decoding step.

Since the revision steps of the encoder and decoder states may be different, we refer this technique as asynchronous revision decoding. For non-streaming model, more revisions means more available right context, then gives better results. However, more revision also leads to larger latency. Nevertheless, revision  provides trade-off between latency and accuracy. In other words, the non-streaming model can be used with any right context during inference, which achieves arbitrary latency. Thus, the latency of a ASR model can be arbitrarily changed after training, which we call dynamic latency.

\subsection{Segment Cropping Training}
\label{ssec:segment}
When training a non-streaming ASR model, the full utterance is provided. At each frame, the encoder has a complete context. When performing asynchronous revision during inference, the finalized encoder states may get incomplete right context. As shown in figure \ref{fig:async_revise}, suppose the right context covers five chunks, the finalized encoder state E0 only sees two chunks in the right since the encoder revision is two steps. Then, E0 provides partial memory to the next state E1 and so on. This leads to inference and training mismatch. 

\begin{figure}[htb!]
  \centering
  \centerline{\includegraphics[width=0.8\columnwidth]{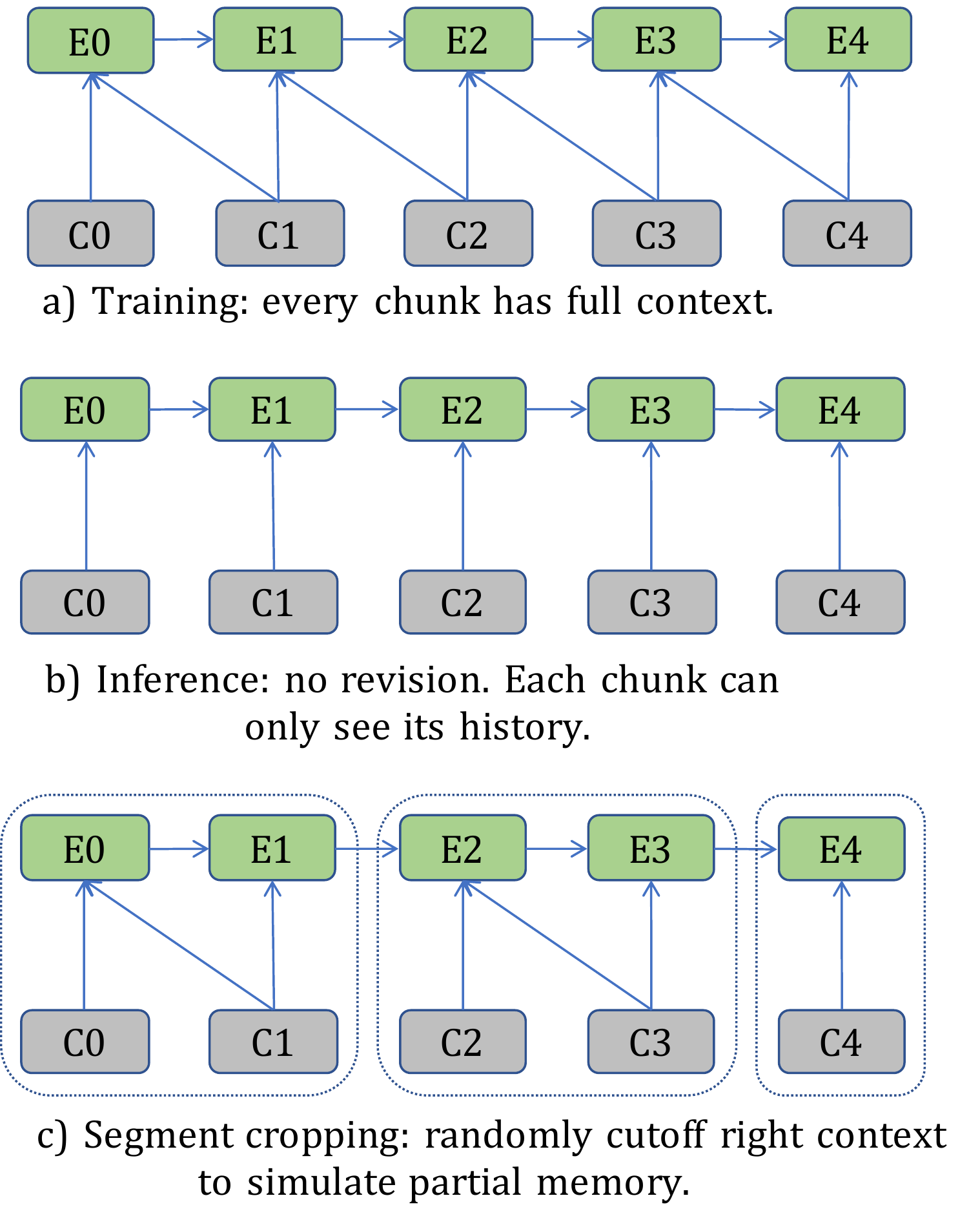}}
\caption{Partial memory problem and segment cropping training. The left context of the encoder states are omitted.}
\label{fig:segment_crop}
\end{figure}

Prefix training is used to alleviate partial inputs in simultaneous (speech) translation\cite{arivazhagan2020retranslation_spoken,arivazhagan2020retranslation}. However, its not work for RNN-T models for its intrinsic monotonic alignment property. In our framework, revision is applied on each chunk during inference, if the right context of the finalized encoder state is not the same as training, mismatch happens. We call this incomplete right context state as partial memory. To alleviate partial memory problem in incremental decoding, we propose a training strategy, segment cropping, which randomly splits input sequence into several segments with forward connections. As shown in figure \ref{fig:segment_crop}, every chunk has full right context in normal training. At inference, partial memory occurs in every chunk if no revision. By segment cropping, we can simulate partial memory in training.

\section{experiments}
\label{sec:exp}


\subsection{Data}
\label{ssec:data}
For our experiments we use 10K hours of internal Mandarin speech dataset. The test set we use consists of 21k utterances with duration less than 30 seconds long from various applications. The training and test sets are all anonymized and hand-transcribed. The input speech wave-forms are framed using a 25 msec window with 10 msec shift. We use 80 dimension filter bank features.  We report the model performance using Character Error Rate (CER).

\subsection{Model Structures and Hyperparameters}
\label{ssec:model_param}
Without loss of generality, our encoder network is based on a temporal convolutional network, DFSMN \cite{zhang2018dfsmn}, which can be extended to self-attentional network like conformer \cite{gulati2020conformer}. And our decoder network is consists of two LSTM layers with 1024 units. At the front of the encoder network, we use two convolution subsampling layers, each with a stride 2. For non-streaming model, the encoder has 30L DFSMNs with 2048 hidden units, the input context for each convolution layer is $[20, 20]$, which means 20 left frames and 20 right frames along with the current frame.  For streaming models, we use three different context configurations: $[20,1] \times 10 + [20,0] \times 20$ for 0.4 second (0.4s for short) latency model, $[20, 1] \times 30$ for 1.2s latency model and $[20, 2] \times 30$ for 2.4s latency model. The only difference for those models is the right context configuration. For simplicity, we use greedy search to decode our models.

\subsection{Asynchronous Revision on Non-Streaming Model}
\label{ssec:async_revise_nonstream}
We perform asynchronous revision decoding on the baseline non-streaming model (CER 9.45\%) with various revisions. At inference,  we divide the input utterances into fixed-size chunks with 40 frames and decode every time a new chunk arrives.  Hence, the number of revisions means the number of history chunks to be revised.

\begin{figure}[htb!]
  \centering
  \centerline{\includegraphics[width=1.0\columnwidth]{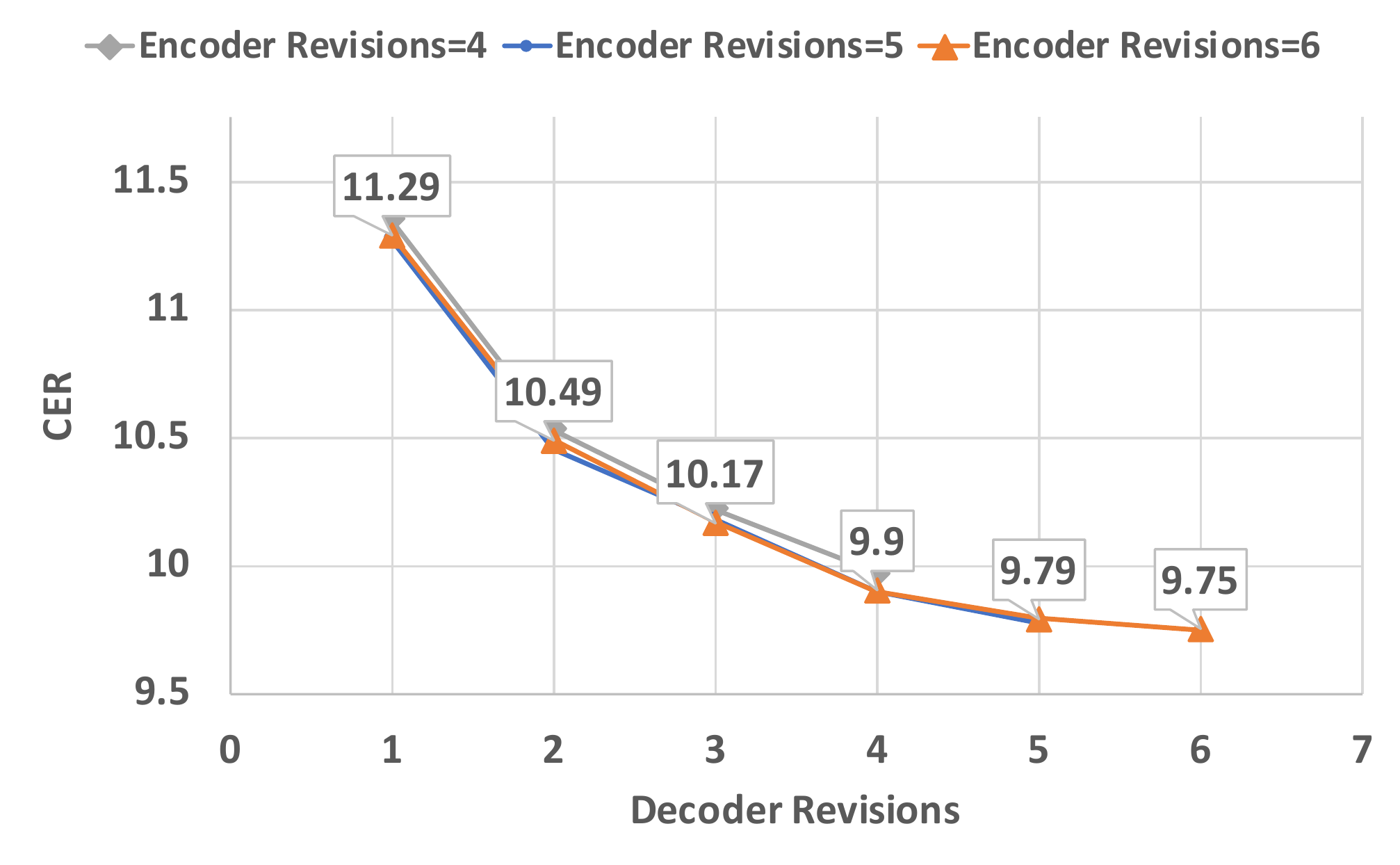}}
\caption{Asynchronous revision on non-streaming model. }
\label{fig:exp1_async_revise}
\end{figure}

Figure~\ref{fig:exp1_async_revise} shows the relationship between revisions and CER.  With the same decoder revisions, more encoder revisions has little impact on model performance. As we can see, more decoder revisions brings better results,  while the marginal improvements becomes smaller. 

We have also done experiments with smaller encoder revision, the results are similar with figure~\ref{fig:exp1_async_revise}. Besides, when there is no decoder revision (revise=0), the decoder network sees partial memory without right context, which leads to bad performance (CER $\sim 40\%$). We will not report the results with no revision in the following experiments.

\subsection{Asynchronous Revision on Streaming Models}
\label{ssec:async_revise_stream}
\begin{figure}[htb!]
  \centering
  \centerline{\includegraphics[width=1.0\columnwidth]{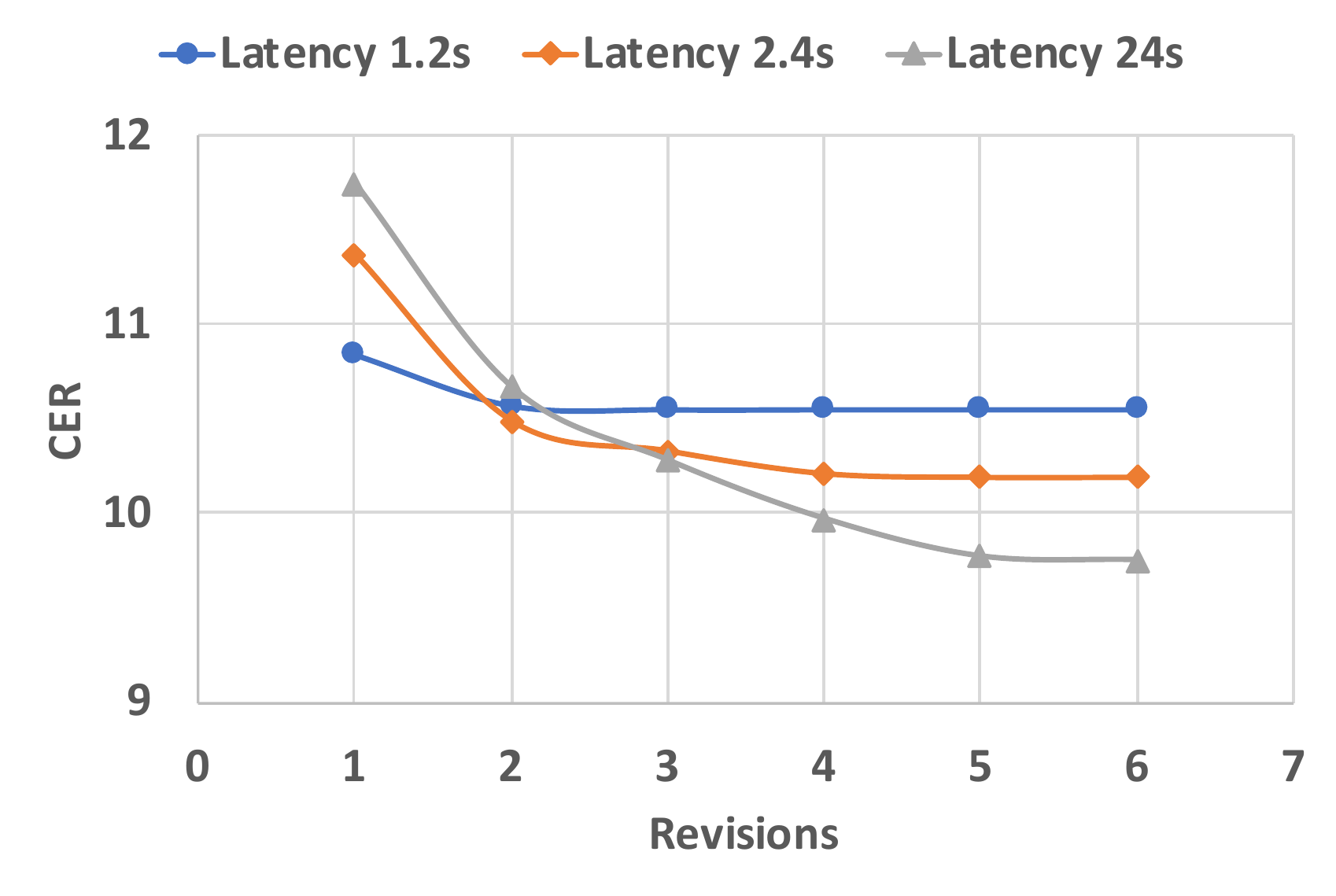}}
\caption{Asynchronous revision on streaming models.}
\label{fig:exp2_decoder_revise}
\end{figure}

The proposed framework can also works on streaming model. As shown in figure~\ref{fig:exp2_decoder_revise}, we perform asynchronous revision on two different streaming models, along with the baseline non-streaming model. For simplicity, we use the same revisions for the encoder and decoder networks. It is evident that the number of revisions is positively related to performance whenever the model latency is. Meanwhile, large revisions gives little relative improvements since the right context provided by revision is close to training. Interestingly, when revisions is small (e.g. revise=1), lower latency model gives a better result. We believe that it is because partial memory has larger impact on higher latency model.

\subsection{Dynamic Latency Versus Streaming}
\label{ssec:dynamic_latency}

\begin{table}[t]
  \centering
  \begin{tabular}{cccc}
    \toprule
    \multicolumn{1}{c}{\textbf{Model}} &
    \multicolumn{1}{c}{\textbf{Revisions}} &
    \multicolumn{1}{c}{\textbf{Latency}}  &
    \multicolumn{1}{c}{\textbf{CER}} \\
    \midrule
    \multirow{3}*{Streaming} & - & 0.4s & 11.63 \\
                            & - & 1.2s & 10.63 \\
                            & - & 2.4s & 10.20 \\
    \midrule
    \multirow{4}*{Dynamic Latency} & 1 & 0.4s & 11.75 \\
                            & 3 & 1.2s & 10.28 \\
                            & 6 & 2.4s & 9.75 \\
                            & - & 24s & 9.45 \\
    \midrule
                            & 1 & 0.4s & \textbf{10.29} \\
    Dynamic Latency  & 3 & 1.2s & \textbf{9.55} \\
    +3 Segments Cropping  & 6 & 2.4s & \textbf{9.32} \\
                            & - & 24s & 9.26 \\
    \midrule
                            & 1 & 0.4s & \textbf{10.03} \\
    Dynamic Latency  & 3 & 1.2s & \textbf{9.55} \\
    +5 Segments Cropping  & 6 & 2.4s & \textbf{9.37} \\
                            & - & 24s & 9.29 \\
    \midrule
  \end{tabular}
  \caption{Dynamic latency model versus streaming models. '-' means full context decoding.}
  \label{tab:dynamic_latency}
\end{table}

As discussed in section~\ref{ssec:async}, the latency of a ASR model can be arbitrarily changed with asynchronous revision during inference. Without loss of generality, the decoding chunk size is 40 frames. Revise one chunk means 0.4 second latency, 3 chunks for 1.2 seconds latency and 6 chunks for 2.4 seconds latency.  We denote the non-streaming model applied asynchronous revision during inference as dynamic latency model.

Table~\ref{tab:dynamic_latency} presents the results of dynamic latency model versus streaming models. The streaming models we used are described in section~\ref{ssec:model_param}. The dynamic latency model gives better or similar results compared with streaming models in the same latency configuration. To alleviate partial memory mismatch, we apply segment cropping training. We use two cropping strategies by randomly splitting input utterances into 3 or 5 segments. As shown in the last two blocks in table~\ref{tab:dynamic_latency}, both strategies give similar results. Compare with streaming models, the dynamic latency model with segment cropping training gives 13.76\%, 10.16\% and 8.14\% relative improvements in 0.4s, 1.2s and 2.4s latency respectively. Furthermore, the performance of the dynamic latency model is extremely close (less than 1\% rel.) to that of the non-streaming model in the case of 2.4 second latency.


\section{conclusion}
\label{sec:conclusion}
We propose an inference framework, asynchronous revision, to perform streaming and non-streaming speech recognition in one model. Moreover, in streaming mode, the latency of a non-streaming model can be arbitrary changed, which we call dynamic latency. Thus can be generally applied as an inference technique without requiring extra training support. We also propose segment cropping training to alleviate partial memory problem during inference.  We show that the dynamic latency model gives 8\%-14\% relative improvements over the streaming models under the same latency. We also show that in the 2.4 seconds latency setting, the performance of asynchronous revision decoding is very close to full context decoding. A limitation of our framework is that more revisions means more computations. Our future work will focus on how to improve the computation efficiency of asynchronous revision.





\newpage

\bibliographystyle{IEEEbib}
\bibliography{main}

\end{document}